# Predictability Measures for Power Quality Time Series in Medium-Term Forecasting

Peter Feistel, Max Domagk, *Member IEEE*, Jan Meyer, *Senior Member IEEE* and Marco Lindner

***Abstract*-- Medium-term forecasting of Power Quality (PQ) parameters, on horizons of weeks to about one year, supports proactive maintenance and the early detection of limit exceedances in transmission network monitoring. Its practical value, however, depends on knowing in advance which time series can be forecast reliably at all. This article addresses that question in two stages, based on 2,807 weekly time series of PQ parameters from a long-term measurement campaign at 66 sites in the German transmission system, covering the 110 kV, 220 kV, and 380 kV levels. First, eight forecasting models are benchmarked. The STL-ARIMA hybrid achieves the highest accuracy, with an average sMAPE of 17.63% and a lower error than the seasonal naive benchmark for 72% of the time series, while accuracy varies substantially across PQ parameters and measurement sites. Second, model-free features computed from the training set alone are evaluated as measures of intrinsic predictability. SVD entropy and the Crest Factor correlate strongly with the realized forecast accuracy of all eight models and are used in a logistic regression to estimate the probability of a poor forecast before any model is applied. The resulting measures allow operators to separate time series suitable for automated forecasting from those requiring manual review.**

## I. Introduction

THE increasing presence of harmonic sources, such as wind farms, solar parks, HVDC systems, and static VAR compensators, highlights the growing importance of Power Quality (PQ) in both transmission and distribution networks [1]. As a result, harmonic measurements in high and extra-high voltage systems have become more important, leading transmission system operators (TSOs) to expand their PQ monitoring efforts aimed at evaluating compliance with emission limits and planning levels such as IEC 61000 standards [2]. Accurate medium-term forecasting of PQ parameters enables early detection of undesirable developments and possible threshold exceedances, supporting proactive maintenance and strategic investment decisions [3].

Forecasts are commonly classified by their horizon into very short-term (minutes to one hour), short-term (hours to a few days), medium-term (weeks to about one year), and long-term (one year to several decades) [4]. This article focuses on medium-term forecasting, which matches the time scale on which planning levels are assessed and network investments are prepared. Long-term forecasting of PQ parameters is additionally constrained by network changes that cannot be anticipated from measurement data alone.

There is limited research on medium-term forecasting of PQ parameters at the transmission level. In both general power system forecasting and PQ-specific applications, most studies have concentrated on short-term horizons, as these are most critical for real-time operation, market scheduling, and risk management. A typical application is load forecasting for operative reserve [4]. Within the context of PQ forecasting, the literature predominantly addresses short-term, site-specific prediction and mitigation of harmonic emissions, often based on relatively small data sets and targeting compliance at individual facilities or feeders [5], [6]. Studies that consider entire networks typically remain focused on the distribution level and short-term horizons, frequently aiming to predict specific PQ disturbances such as harmonic distortion or voltage dips [7]-[11].

Reference [12] addressed harmonic distortion forecasting in sparsely monitored transmission networks with significant renewable generation, focusing on short-term horizons to enhance visibility at unmonitored locations. While their approach improves spatial coverage, it does not target medium- or long-term trend forecasting. To the best knowledge of the authors, no comprehensive framework currently exists that integrates medium- or long-term forecasting into continuous PQ monitoring at the transmission level. The present work adopts the operator's perspective, aiming to provide site-level forecasts that extend the temporal scope of existing PQ monitoring practices.

Beyond accuracy, the reliability of a forecast should be anticipated, since every model has inherent limitations and it is valuable to estimate in advance whether a given TS is predictable at all. Prior studies on load data have introduced indicators of short-term predictability, revealing systematic differences in forecastability across datasets [13]. Recent work on photovoltaic generation shows that entropy-based measures can quantify short-term predictability, yet measures for medium-term predictability remain scarce despite their relevance for planning and risk assessment [14].

This article addresses these challenges through a statistical analysis of PQ data from a long-term measurement campaign, following a two-stage approach. In the first stage, eight

P. Feistel, M. Domagk, and J. Meyer are with TUD Dresden University of Technology, Dresden, Germany (email: max.domagk@tu-dresden.de)
M. Lindner is with TransnetBW GmbH, Stuttgart, Germany

forecasting models are benchmarked on 2,807 PQ time series in order to identify the modeling paradigms suited to medium-term forecasting and to establish the realized predictability of each time series. In the second stage, model-free features computed from the training set alone are evaluated as measures of intrinsic predictability, and the most informative ones are used to estimate, before any model is applied, whether a reliable forecast can be expected. Both stages together allow operators to balance automated forecasting against manual supervision within existing PQ monitoring frameworks.

The remainder of this article is organized as follows. Section II introduces the concept of predictability and the distinction between its intrinsic and realized form. Section III describes the measurement dataset and the preprocessing procedure. Section IV benchmarks the forecasting models and identifies those best suited to medium-term forecasting of PQ parameters. Section V evaluates time series features as measures of intrinsic predictability and applies the most informative ones to the estimation of forecast reliability.

## II. Predictability of PQ Time Series

### A. Concept of Predictability

A time series (TS) can be regarded as a realization of an underlying stochastic or dynamic system, that is, a system whose state evolves over time according to deterministic rules, random influences, or a combination of both [15]. Predictability describes to what extent the future evolution of such a system can be inferred from past observations. Two fundamentally different notions have to be distinguished [16], [17].

Intrinsic predictability $\rho_\mathrm{I}$ is a property of the data itself and is independent of the forecasting model. It represents the theoretical upper bound of the achievable forecast accuracy and is derived from the available past observations alone, formally $\rho_\mathrm{I} = \rho(S)$ for a time series $S$ [15].

Realized predictability $\rho_\mathrm{R}$ is model-dependent and is quantified by accuracy measures evaluated on observations that were not used to fit the model, formally $\rho_\mathrm{R} = \rho(M, S)$ for a model $M$ applied to $S$ [18]. It reflects the performance of one particular model rather than a property of the time series, since different models may attain different levels of accuracy for the same time series because they rest on different structural assumptions.

The distinction is necessary because a large forecast error does not discriminate between an inadequate model and a time series that is inherently difficult to predict [17]. Forecast accuracy therefore quantifies realized, not intrinsic, predictability [16]. Throughout this article, predictability is used in the sense defined here; the term forecastability used in [13] is treated as equivalent.

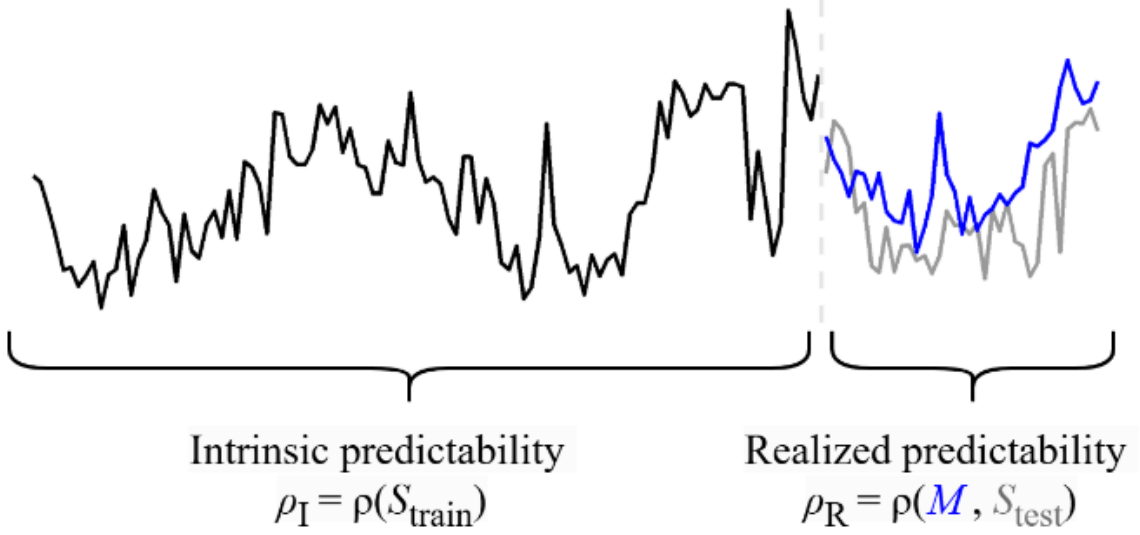

Fig. 1. Illustration of concept of predictability for a time series using the training set ($S_\mathrm{train}$), test set ($S_\mathrm{test}$) and the forecast of a model ($M$)

Fig. 1 illustrates the estimation of both quantities from a time series split into a training set ($S_\mathrm{train}$) and a test set ($S_\mathrm{test}$). Intrinsic predictability $\rho_\mathrm{I}$ is estimated from $S_\mathrm{train}$, while realized predictability $\rho_\mathrm{R}$ follows from comparing the forecast of the model $M$, fitted on $S_\mathrm{train}$, with $S_\mathrm{test}$.

### B. Realized Predictability

The forecast error, defined as the difference between the observed value $y_{T+h}$ and the predicted value $\hat{y}_{T+h}$ for a horizon $h$:

$$e_{T+h} = y_{T+h} - \hat{y}_{T+h} \tag{1}$$

forms the basis of most accuracy measures, where large errors imply low realized predictability. Such measures are well suited for comparing forecasting models, but they do not reveal the maximum accuracy attainable for a given time series under an optimal model. For PQ time series, realized predictability is obtained by comparing model forecasts with the corresponding test set over the given horizon. The accuracy measures applied in this article are defined in Section IV.B.

### C. Intrinsic Predictability

Model-free estimates of intrinsic predictability are mostly based on information-theoretic measures [17], which quantify the degree of irregularity, that is, the absence of recurring patterns in the data sequence [19]. Since these measures operate on the data sequence itself rather than on a fitted model, they also capture regularity that is not apparent to visual inspection. For PQ time series, intrinsic predictability is estimated from the training set alone and can therefore be evaluated before any model is applied, in order to determine whether a reliable forecast can be expected for a particular time series. The features examined for this purpose are introduced in Section V.A.

## III. Data and Preprocessing

### A. Power Quality Parameter Variations

PQ parameters vary over time due to the influence of many different factors, mainly related to the electrical environment. Generally, three different types of variations can be distinguished: short-, medium- and long-term variations [3]. Short-term variations are usually characterized by daily and weekly intervals which are mainly influenced by the consumer and generation behavior. Certain consumer configurations show distinctly different daily patterns (e.g., working days vs. weekends).

Medium-term variations over periods of several months are linked to more general changes in consumer and generation behavior that are mostly driven by seasonal effects. Earlier nightfall in winter months can lead to an increased use of equipment such as lighting, and consequently to higher emission of harmonics. The generated energy of PV systems is significantly higher during the summer months (longer days, higher angle of insolation) [20].

Long-term variations extend over several years and originate from structural changes, such as newly connected generation or load, modifications of the network topology, or the gradual replacement of equipment technology.

This article focuses on medium-term variations. They dominate on horizons of weeks to about one year and provide the seasonal structure that the forecasting models exploit, whereas long-term variations are driven by network changes that cannot be anticipated from measurement data alone.

### *B. Data Set*

The dataset originates from a measurement campaign within the German transmission system. It contains 66 measurement sites spanning three different voltage levels: 110 kV (30 sites), 220 kV (14 sites), and 380 kV (22 sites). The duration of the measurements ranges from at least 2.5 years up to 4 years for all recorded data. Data acquisition employed fixed PQ instruments that comply with the IEC 61000-4-30 Class A standard.

The monitored PQ parameters include voltage quality parameters: long-term voltage flicker ($U_{\mathrm{Plt}}$), voltage unbalance ($u_2$), total harmonic distortion of voltage ($U_{\mathrm{THD}}$) and individual, most prominent harmonic voltages ($U^{(h)}$) of order $h \in \{3, 5, 7, \dots, 15\}$. On the current side, the total harmonic current ($I_{\mathrm{THC}}$) and the corresponding individual harmonic currents $I^{(h)}$ of the same orders are monitored, each representing either a whole substation or a specific feeder within the network. In addition to the PQ parameters, the total RMS current ($I_{\mathrm{RMS}}$) recorded by the same instruments is included in the analysis. It characterizes the loading of the measurement point, and its medium-term development is of direct interest for network planning. It is therefore treated as a further time series in the benchmark.

### *C. Data Preprocessing*

*1) Uncertainty assessment*

The amplitudes for the harmonic voltages and currents usually reduce with higher harmonic orders. Especially for triplen harmonic orders (e.g. $h = 3$, 9, and 15) the measured values are often close to the instrument's measurement noise floor. To ensure sufficient data quality, an uncertainty threshold is applied. Measurements falling below this threshold are considered to exhibit an unacceptably high error. This threshold was determined in a laboratory test for the specific type of PQ instrument used. The following uncertainty thresholds have been derived for a maximum amplitude error of 10% for harmonic voltages and currents [3]:

- Harmonic voltages: 14 mV (instrument input)
- Harmonic currents: 0.33 mA (instrument input).

*2) Calculation of weekly 95$^{th}$ percentiles*

The weekly 95th percentiles are calculated for every full calendar week (from Monday to Sunday). A week is considered valid if at least 95% of all measured values are available. This criterion allows up to 8.5 hours of missing data per week, in order to increase the amount of available measurement data.

*3) Impute missing weeks*

Missing weeks, or gaps within TS, frequently occur, often due to interruptions such as maintenance work. To maximize the number of available time series for analysis at most 20% missing weeks and at most 10 consecutive missing weeks are permitted [3]. Missing weeks are imputed with the last calculated weekly percentile before the gap.

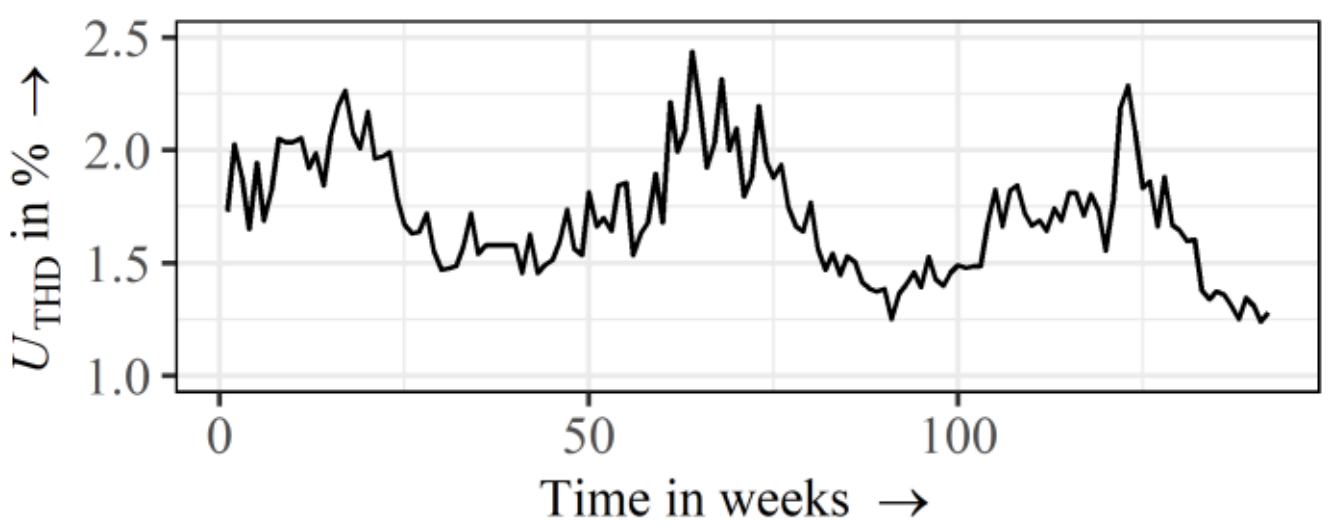

Fig. 2. Example time series of the total harmonic voltage distortion ($U_{\mathrm{THD}}$) with a measurement duration of 141 weeks at the 110-kV voltage level

*4) Select suitable time series of weekly percentiles*

A minimum measurement duration of 141 weeks was chosen to maximize the number of included time series. After preprocessing, 2,807 time series are available for analysis.

An example time series for the total harmonic voltage distortion is illustrated in Fig. 2. This example demonstrates the typical characteristics observed in the data, including seasonal variations (with higher emission levels during summer months) and a slightly decreasing trend.

## IV. Realized Predictability of PQ Time Series

### *A. Forecasting Models*

While machine and deep learning models have excelled in short-term forecasting, their practical advantage for medium- and long-term horizons is often limited, supporting the hypothesis that long-term feasibility relies on distinct trend and periodicity structures [21]. Furthermore, these approaches often act as black boxes, lacking explainability and demanding significant computational resources [22]. For these reasons, this contribution adopts models that specifically treat time series as a sum of interpretable components (trend, seasonal, and remainder).

*1) Baseline Models*

The Naive approach projects all future values as the last observed data point. The Drift model is a variant that extrapolates by the average change (drift) observed between the first and last observations. For TS exhibiting strong seasonality, the Seasonal Naive (SNaive) approach is employed. It is based on the principle that future values will equal observations from the corresponding seasonal period in the previous cycle. For instance, in PQ data with yearly periodicity, the forecast for a given week repeats the observation from the same week one year earlier. SNaive is commonly utilized as a persistence model and serves as the essential baseline for benchmarking forecast accuracy in this study [23].

*2) Exponential Smoothing Models*

Exponential Smoothing (ES), originating from the work of [24]-[26], generates forecasts as exponentially weighted averages of past observations. Its extensions, such as Holt's and Holt-Winters' model, incorporate trend and seasonal components and are particularly suitable for stable seasonal

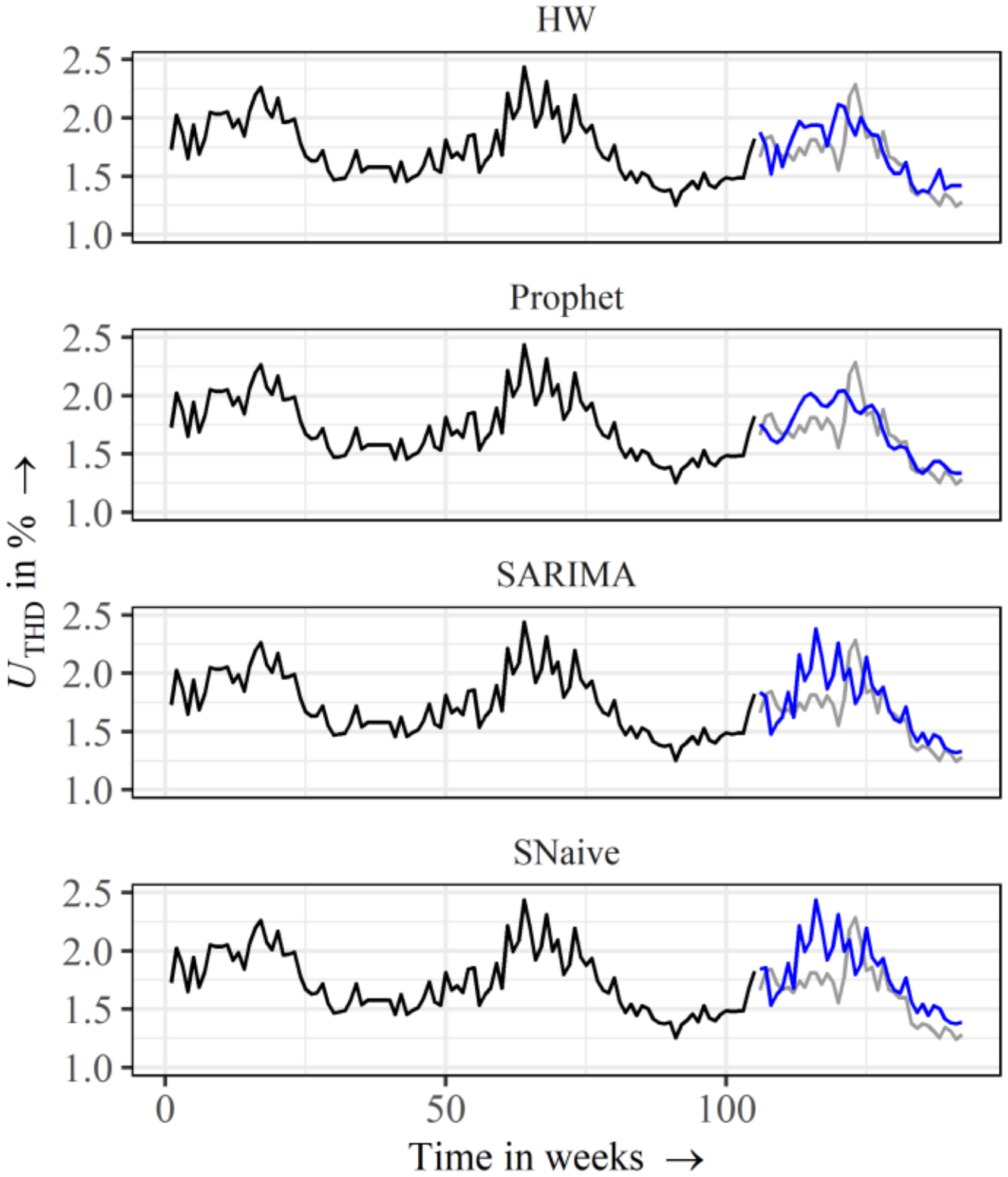

Fig. 3. Example forecast of the total harmonic voltage distortion ($U_{THD}$) using four different models; training set (105 weeks) in *black*; test set in *grey* (36-week horizon) and model forecasts in *blue*.

patterns.

*3) ARIMA Models*

The Box-Jenkins methodology models autocorrelation structures in the data and can incorporate both non-seasonal (ARIMA) and seasonal (SARIMA) differencing [27]. These models are particularly effective for time series with strong autocorrelations and well-defined seasonal cycles.

*4) Prophet Model*

The Prophet model [28] is an additive regression framework with components for trend, seasonality, and holiday effects. It supports flexible trend modeling, specifically using piecewise-linear or logistic growth curves, and captures multiple seasonalities via Fourier series, offering flexibility for data with both periodic and non-periodic variations.

*5) STL Decomposition and Hybrid Models*

In practice, STL decomposition (Seasonal and Trend decomposition using Loess) [29], [30] can be applied prior to modeling, enabling different forecasting models to be applied to separate components. For instance, the SNaive model may be used exclusively for the seasonal component, while another model like ARIMA is applied to the seasonally adjusted series, as demonstrated in [31].

Fig. 3 shows example forecasting results for a time series of the total harmonic voltage distortion ($U_{THD}$), obtained using four different forecasting models.

## B. Assessment Procedure

Typical measures for point-forecast accuracy are Mean Absolute Error (MAE) and Root Mean Squared Error (RMSE). These metrics are scale-dependent, which makes them unsuitable for comparing TS across different locations and PQ parameters with changing units and value ranges. To account for differences in scale, a normalized RMSE (nRMSE) can be used, where the RMSE is divided by a characteristic value such as the range of the observed data [32].

The Mean Absolute Percentage Error (MAPE) corresponds to the mean absolute deviation between predicted and observed values and is independent of the value range. Nevertheless, MAPE is asymmetric, weighting positive errors (predicted values smaller than observed values) higher than negative errors. Negative errors can reach a maximum of 100% while positive errors can become infinitely large [33].

To address this issue, symmetric Mean Absolute Percentage Error (sMAPE) was introduced to balance the asymmetry of MAPE [34]:

$$\text{sMAPE} = \frac{200}{H}\sum_{h=1}^{H}\frac{|y_{T+h}-\hat{y}_{T+h}|}{|y_{T+h}|+|\hat{y}_{T+h}|}. \tag{2}$$

Although this measure is "symmetric" in the range [0%, 200%], it becomes unstable for small values near zero [35]. As an alternative, the Mean Absolute Scaled Error (MASE) can be used [32]. It scales the forecast errors by the in-sample mean absolute error of a naive benchmark, allowing direct comparison across series of different scales and forecast horizons. Based on these measures, the following procedure is used to compare the candidate models.

The candidate forecasting models introduced in Section IV.A are evaluated on all 2,807 TS. Because many PQ time series exhibit strong annual seasonality, STL decomposition is combined with several of these models. After decomposition into a seasonal component, a trend component, and a remainder, a seasonal naive forecast is applied to the seasonal component, while four models, namely Drift, Exponential Smoothing, Holt's model, and ARIMA, are applied to the seasonally adjusted series. Together with SNaive, Holt-Winters, SARIMA, and Prophet, this yields the eight models listed in Tab. I, with SNaive serving as the benchmark throughout.

To ensure consistency and comparability, a training period of two years (105 weeks) is used for each TS, and the forecast horizon is set to 36 weeks.

TABLE I

Overview of forecasting models evaluated in this study

| Model | Description |
|---|---|
| SNaive | Seasonal Naive (benchmark) |
| HW | Holt-Winters model |
| SARIMA | Seasonal ARIMA |
| Prophet | Additive model with trend and seasonality |
| STL-Drift | STL Decomposition with Drift model |
| STL-ES | STL Decomposition with Simple Exponential Smoothing |
| STL-Holt | STL Decomposition with Holt's model |
| STL-ARIMA | STL Decomposition with ARIMA model |

The procedure for assessing the forecasting accuracy of the different models follows the approach outlined in [36] and consists of three steps:

1. Calculation of the sMAPE accuracy measure over the entire forecast horizon for each model and each TS.
2. Determination of the rank R(sMAPE) for each model and each TS.
3. Calculation of the mean accuracy measure $\overline{\text{sMAPE}}$ and mean rank $\overline{\text{R}}$ for each model across all TS.

The main advantage of using mean ranks is their greater robustness to outliers compared to the mean values of the accuracy measures. A Benchmark Ratio is additionally computed to quantify each model's performance relative to the SNaive baseline:

$$\text{BR}_i = \overline{\text{sMAPE}}_i / \overline{\text{sMAPE}}_{\text{SNaive}}. \tag{3}$$

### C. Identification of Best Performing Forecasting Models

The individual models were applied to the data set of 2,807 TS. The average forecast accuracy and ranks, shown in Tab. II, were calculated and sorted by increasing average sMAPE, with SNaive serving as a baseline.

TABLE II

Averages of sMAPE and corresponding ranks for 2,807 time series across different forecasting models (training: 105 weeks; test: 36 weeks; WR: win rate against SNaive)

| Rank | Model | $\overline{\text{sMAPE}}$ | $\overline{\text{R}}$ | BR | WR |
|---|---|---|---|---|---|
| 1 | STL-ARIMA | 17.63% | 3.4 | 0.936 | 72.3% |
| 2 | STL-ES | 18.44% | 3.59 | 0.980 | 65.9% |
| 3 | SNaive (benchmark) | 18.83% | 4.93 | 1 | - |
| 4 | SARIMA | 18.87% | 4.27 | 1.002 | 57.9% |
| 5 | Prophet | 19.19% | 4.13 | 1.019 | 57.4% |
| 6 | STL-Holt | 20.37% | 4.51 | 1.082 | 55.0% |
| 7 | STL-Drift | 22.77% | 5.29 | 1.209 | 46.0% |
| 8 | HW | 23.89% | 5.88 | 1.269 | 37.8% |

As shown, the STL-ARIMA model performs best with respect to both measures, followed by STL-ES. A notable observation is that, although SNaive achieves a better average sMAPE than three of the other seven models, it exhibits poorer rank performance, ranking below six of the seven models in terms of R(sMAPE). The win rate (WR) against SNaive, defined as the share of TS for which a model achieves a lower sMAPE than SNaive, further confirms this: STL-ARIMA and STL-ES outperform SNaive on a per-series basis in 72% and 66% of cases, respectively.

In addition, Fig. 4 shows the distributions of the sMAPE across all TS.

STL-ARIMA stands out as the best model. Five of the seven

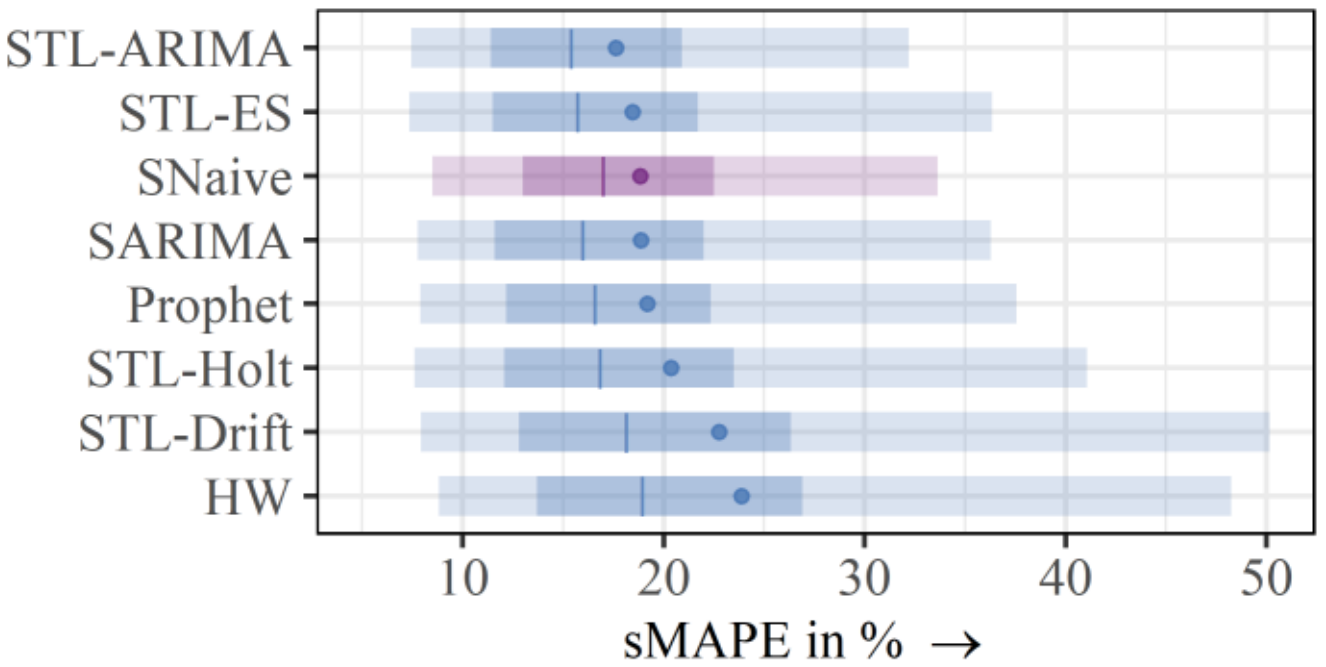

Fig. 4. Distribution of sMAPE for 2,807 time series across different models (*blue*) ordered by mean (*dot*); median (*line*); 25–75% range (*dark bar*); 5–95% range (*light bar*); SNaive benchmark (*purple*)

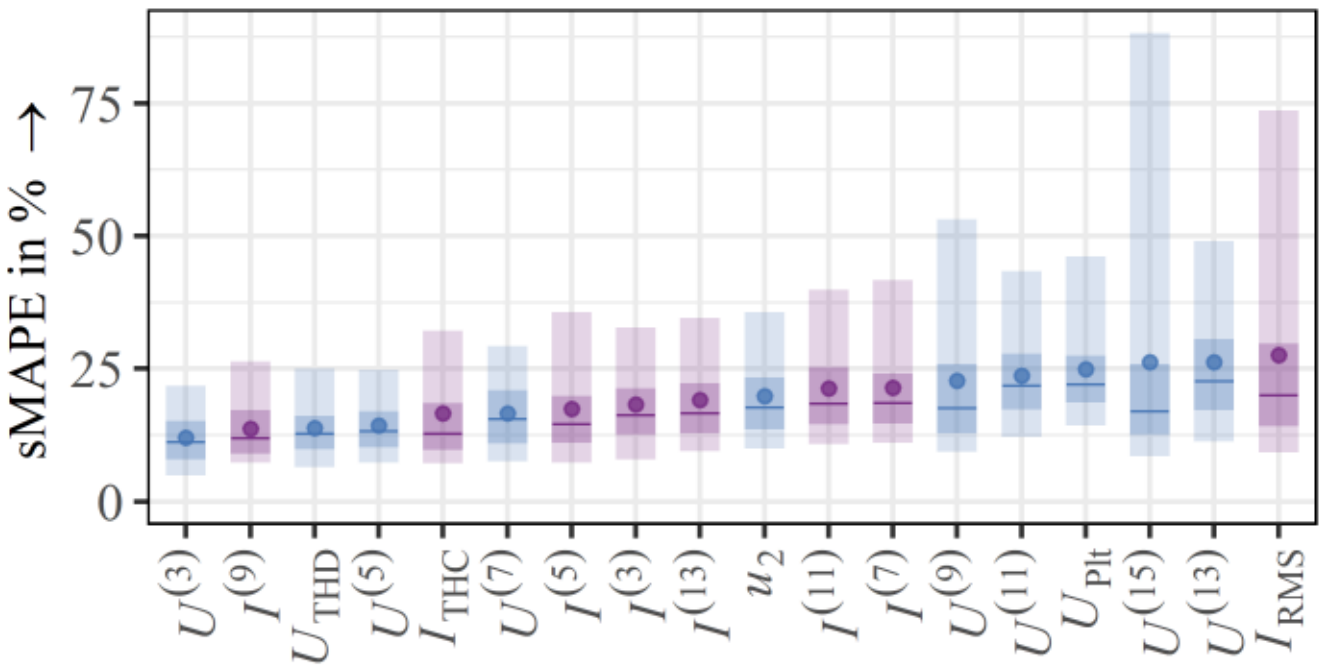

Fig. 5. Distribution of sMAPE for 2,807 time series across different current (*purple*) and voltage (*blue*) quality parameters ordered by mean (*dot*); median (*line*); 25–75% range (*dark bar*); 5–95% range (*light bar*)

models outperform SNaive in terms of the median; however, the 75% quantiles of HW, STL-Holt and STL-Drift are higher than those of SNaive. Furthermore, the 95% quantiles of all models except those of STL-ARIMA exceed those of SNaive. The poor performance of STL-Drift and HW is also very evident from the 95% quantiles with sMAPE > 45%.

Fig. 5 shows clear differences in forecast accuracy across PQ parameters. Lower-order harmonics (3, 5, 7), distortion ($U_{\text{THD}}$,

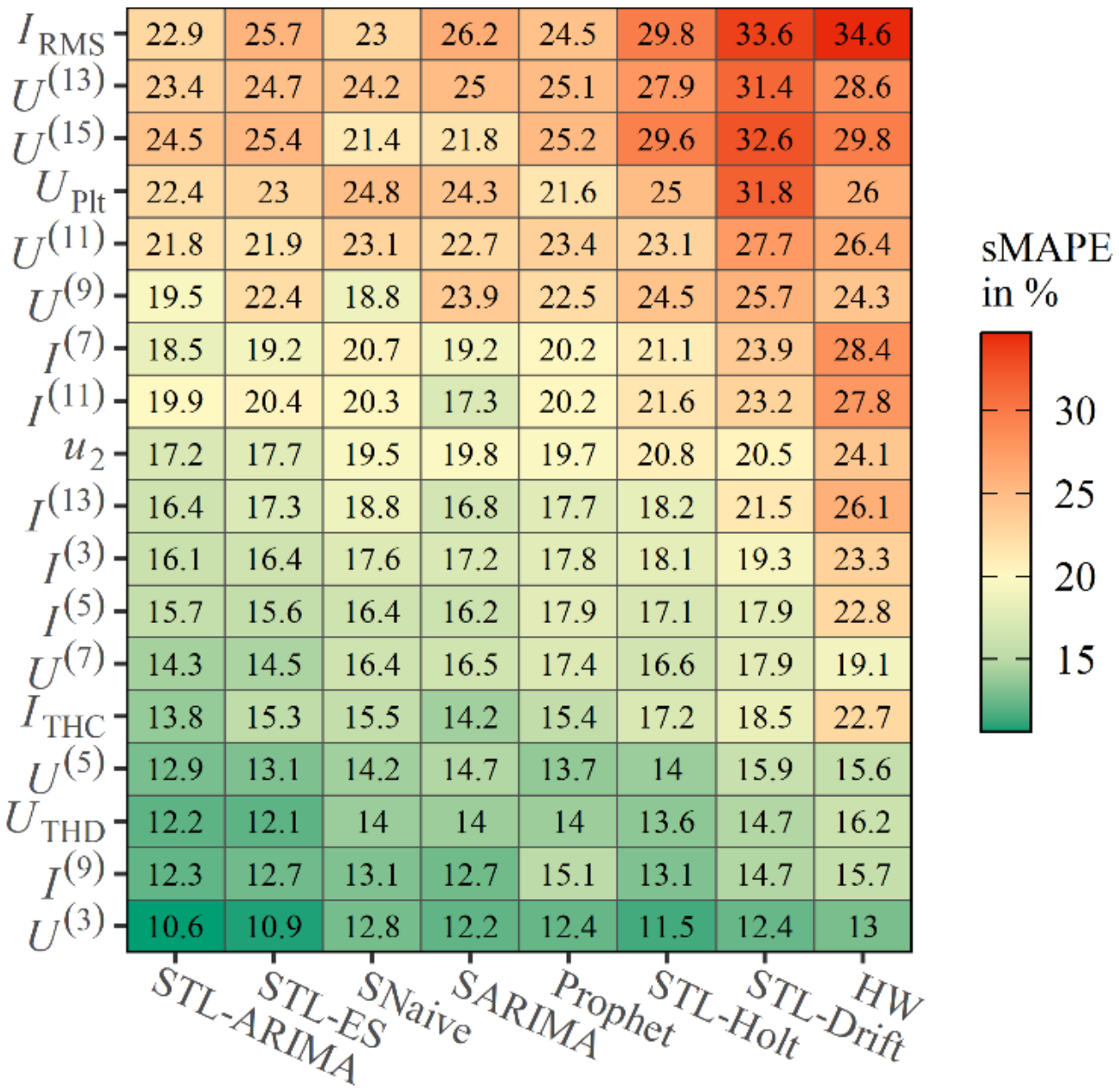

Fig. 6. Average forecast accuracy (sMAPE) across all available time series for each combination of power quality parameters and different forecasting models

$I_{\text{THC}}$), and voltage unbalance ($u_2$) achieve the lowest sMAPE values, while higher-order harmonics (11, 13, 15) and long-term flicker ($U_{\text{Plt}}$) show the poorest accuracy. $U^{(15)}$ and $I_{\text{RMS}}$ also stand out with the highest 95th percentile sMAPE among all parameters.

Fig. 6 compares average sMAPE across models and PQ parameters. STL-ARIMA consistently performs best, with SARIMA approaching its accuracy for higher-order current harmonics (7, 9, 11) and Prophet yielding the lowest sMAPE for $U_{\text{Plt}}$. HW systematically underperforms, and no model surpasses SNaive for $U^{(15)}$, suggesting limited predictability of that parameter. Forecasting performance depends jointly on model choice and TS characteristics. Decomposition-based models benefit from regular, structured TS data, whereas irregular TS data remain difficult to predict regardless of the model. This is evident at five measurement sites where 20-30% of all time series show poor forecast accuracy (sMAPE > 25%) across all models, and for PQ parameters such as $U_{\mathrm{Plt}}$ and higher-order harmonics, where 6-10% of series consistently yield poor forecasts.

This suggests that reduced predictability is primarily driven by intrinsic TS characteristics rather than model choice.

Fig. 7 illustrates two reasons for low predictability. First, randomly occurring (step) changes result from unpredictable on/off switching of a combined heat and power plant according to market conditions (top).

Second, sporadic outliers appear as individual weeks with extreme spikes reaching more than twice the base level, which is a common feature of voltage flicker (bottom) but also observed in higher-order harmonics (e.g. 13, 15).

## V. Intrinsic Predictability of PQ Time Series

This section investigates the relationship between intrinsic predictability features and realized forecasting performance. The central hypothesis is that features quantifying irregularity and spikiness correlate with forecast accuracy, where low feature values indicate high predictability and high values indicate poor predictability.

### A. Time Series Features

The following features are selected to capture two key characteristics associated with reduced predictability: random regime changes, quantified by randomness features, and sporadic spikes, quantified by spikiness features.

#### 1) Randomness Features

Randomness in sequences can be viewed as a lack of regularity or repetition. Many regularity statistics are based on Shannon entropy [19], [37], which measures the uncertainty of a random variable $Y$:

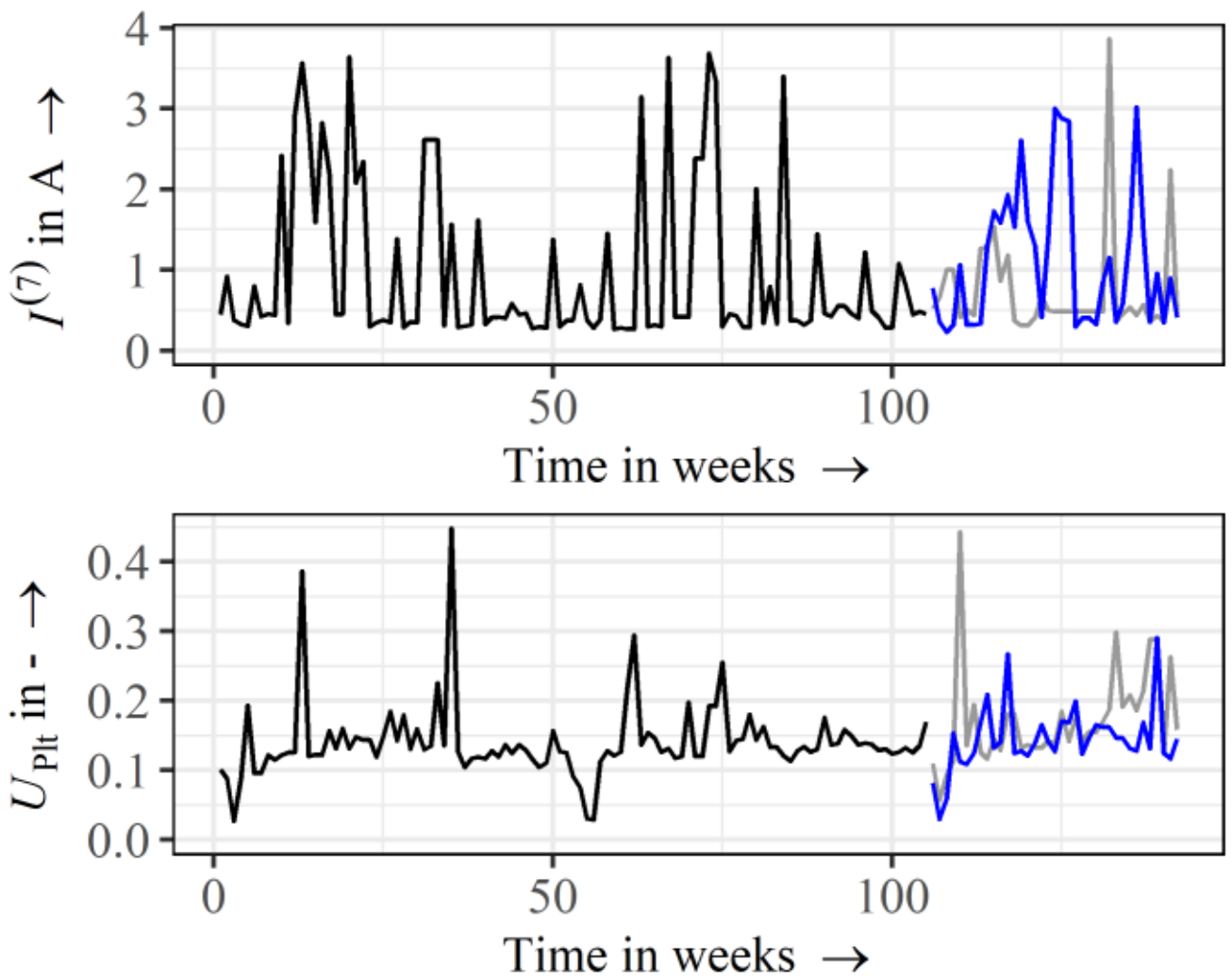

Fig. 7. Representative examples of time series with low realized predictability: the 7th current harmonic exhibiting random regime shifts (*top*) and the long-term flicker with sporadic outliers (*bottom*)

$$H(Y) = -\sum_y p_y \log p_y \quad (4)$$

where $p_y$ is the probability of outcome $y$. Entropy is highest when all outcomes are equally probable and lowest (zero) when the process is fully predictable. For sequential data, however, individual probabilities are typically unknown. Therefore, various measures derived from the TS data itself are used instead, each estimating regularity or complexity in a different way. Six such measures are described below.

*a) Spectral Entropy*

Spectral entropy ($f_{\mathrm{SpE}}$) is computed from normalized power spectral densities $P(f_i)$ over $n$ frequency bins obtained via a discrete Fourier transform [38]:

$$f_{\mathrm{SpE}} = -\frac{1}{\log n}\sum_{i=1}^{n} \tilde{P}(f_i) \log \tilde{P}(f_i) \quad (5)$$

where $\tilde{P}(f_i)$ are normalized spectral powers. The normalization by $\log n$ constrains $f_{\mathrm{SpE}}$ to $[0,1]$. High spectral entropy, indicated by a broad spread of spectral energy, corresponds to low signal regularity. Low spectral entropy, reflecting the dominance of a few frequencies, corresponds to high regularity.

*b) Approximate Entropy*

Approximate entropy ($f_{\mathrm{ApE}}$) quantifies time-series regularity by measuring the logarithmic probability that similar sequences of length $m$ remain similar at the next point $m+1$ [39]:

$$f_{\mathrm{ApE}} = \Phi^m(r) - \Phi^{m+1}(r) \quad (6)$$

with

$$\Phi^m(r) = \frac{1}{N-m+1}\sum_{i=1}^{N-m+1} \log\left(C_i^m(r)\right)$$

where $N$ is the sequence length, $m$ is the embedding dimension (typically 2 or 3), $r$ is the resolution (recommended 0.1-0.25 times the data's standard deviation), and $C_i^m$ is the similarity count. For a highly regular, trending signal, the probability of continued similarity is high, and $f_{\mathrm{ApE}}$ approaches 0. Conversely, more complex, irregular signals yield higher $f_{\mathrm{ApE}}$ values. The calculation includes self-matches, which introduces a statistical bias [19].

*c) Sample Entropy*

Sample entropy ($f_{\mathrm{SaE}}$) is a related regularity measure that improves upon Approximate entropy by excluding self-matches and simplifying the calculation [40]:

$$f_{\mathrm{SaE}} = -\log\left(\frac{\sum_{i=1}^{N-m+1} C_i^{m+1}(r)}{\sum_{i=1}^{N-m+1} C_i^{m}(r)}\right) \quad (7)$$

The parameters $N$, $m$, and $r$ are defined as for Approximate entropy. Both $f_{\mathrm{ApE}}$ and $f_{\mathrm{SaE}}$ are scale-variant; therefore, time series are normalized via z-scoring ($z_t = (y_t - \bar{y})/s_y$) prior to analysis to enable comparisons across different scales. For the following analyses, the embedding dimension is set to $m = 2$.

*d) SVD Entropy*

SVD entropy ($f_{\mathrm{SVDE}}$) is a complexity measure derived from the singular value decomposition (SVD) of an embedding matrix [41]:

$$f_{\mathrm{SVDE}} = \frac{1}{\log(m)} \sum_{i=1}^{m} \left(\frac{\sigma_i}{\sum_{j=1}^{m} \sigma_j}\right) \log\left(\frac{\sigma_i}{\sum_{j=1}^{m} \sigma_j}\right) \tag{8}$$

Here, $\sigma_i$ are the singular values of the embedding matrix $X$, which is constructed from the TS with embedding dimension $m$. A signal with high regularity (e.g., a strong trend) concentrates its energy into a single, dominant singular value, resulting in low $f_{\mathrm{SVDE}}$. The measure is normalized and robust to the choice of $m$, as increasing the dimension primarily adds negligible singular values without altering the normalized energy distribution. For the following analyses, $m = 3$ is applied.

*e) Permutation Entropy*

Permutation entropy ($f_{\mathrm{PE}}$) quantifies time-series complexity by analyzing the relative frequency of ordinal patterns within its subsequences [42]:

$$f_{\mathrm{PE}} = \frac{1}{\log(m!)} \sum_{i=1}^{m!} p(\pi) \log\big(p(\pi)\big) \tag{9}$$

where $m$ is the embedding dimension, $p(\pi)$ is the relative frequency of ordinal pattern $\pi$, and $m!$ is the number of possible patterns. The parameter $m$ is critical, as the number of patterns grows factorially; it is set to $m = 3$ for the following analyses. A weighted variant, $f_{\mathrm{wPE}}$, incorporates the variance of each subsequence, giving greater importance to patterns formed by larger amplitude fluctuations [43].

*2) Spikiness Features*

Spikiness features quantify the presence of sporadic, high-amplitude deviations within a TS.

*a) Crest Factor*

The Crest Factor ($f_{\mathrm{CF}}$) is a common signal analysis metric defined as the ratio of the peak absolute value to the root mean square (RMS) value:

$$f_{\mathrm{CF}} = \frac{\max(|y_t|)}{\sqrt{\frac{1}{N}\sum_{i=1}^{N} y_t^2}} \tag{10}$$

*b) Leave-One-Out Variance*

The Leave-One-Out Variance ($f_{\mathrm{LOOV}}$) quantifies how strongly the variance of a time series changes when each data point is removed in turn [44]. A single point whose removal substantially reduces the variance is identified as a significant outlier. The measure is defined as the mean squared deviation of the leave-one-out variances $s^2_{(-\tau)}$ around their mean $\bar{v}$:

$$f_{\mathrm{LOOV}} = \frac{1}{N-1} \sum_{\tau=1}^{N} \left(s^2_{(-\tau)} - \bar{v}\right)^2 \tag{11}$$

with

$$\bar{v} = \frac{1}{N} \sum_{\tau=1}^{N} s^2_{(-\tau)} \text{ and } s^2_{(-\tau)} = \frac{1}{N-2} \sum_{t=1, t\neq\tau}^{N} \left(y_t - \bar{y}_{(-\tau)}\right)^2$$

where $s^2_{(-\tau)}$ is the variance of the z-scored time series with observation $\tau$ omitted.

*c) Spike Index*

The Spike Index ($f_{\mathrm{SI}}$) is a specialization of $f_{\mathrm{LOOV}}$ applied to the residual component of an STL decomposition of the z-scored time series [44]. By first removing trend and seasonal components, systematic patterns are eliminated and $f_{\mathrm{LOOV}}$ is applied exclusively to the remainder $R_t$:

$$f_{\mathrm{SI}} = f_{\mathrm{LOOV}}(R_t) \tag{12}$$

This isolates sporadic outliers that are independent of the underlying trend level and seasonal structure.

## B. *Assessment Procedure*

Spearman's rank correlation coefficient $r_{\mathrm{S}}$ was used to assess the relationship between intrinsic predictability metrics and realized predictability metrics. Its robustness to outliers and ability to capture monotonic relationships [45] make it preferable to Pearson's correlation, which can be distorted by the extreme values produced by unbounded metrics such as MAPE [46]. The correlation is evaluated against the full set of scale-independent accuracy measures defined in Section IV.B, so that the identification does not depend on a single accuracy measure. The analysis is based on the forecasts of 2,807 time series obtained with a two-year training set and a 36-week horizon. Correlations are computed separately for each forecasting model rather than pooled across model and time series combinations, since pooling would violate the independence of observations.

## C. *Identification of Best Intrinsic Predictability Measures*

Fig. 8 presents the correlations between intrinsic and realized predictability measures. Strong correlations (≥0.76) are observed between the SVD entropy feature ($f_{\mathrm{SVDE}}$) and several accuracy measures, including sMAPE, MAPE, and the normalized MAE and RMSE. The Crest Factor ($f_{\mathrm{CF}}$) also shows notable correlations (≥0.68) with the same measures. In contrast, correlations for both features with MASE and nRMSE are notably weak.

The key finding is that sMAPE demonstrates a strong, consistent correlation with both SVD entropy ($f_{\mathrm{SVDE}}$) and the Crest Factor ($f_{\mathrm{CF}}$), identifying them as prominent indicators of forecasting difficulty. This pattern holds across all evaluated forecasting models, indicating that $f_{\mathrm{SVDE}}$ and $f_{\mathrm{CF}}$ capture intrinsic properties of the time series rather than model-specific behavior, although generality beyond the evaluated model classes cannot be established from this analysis.

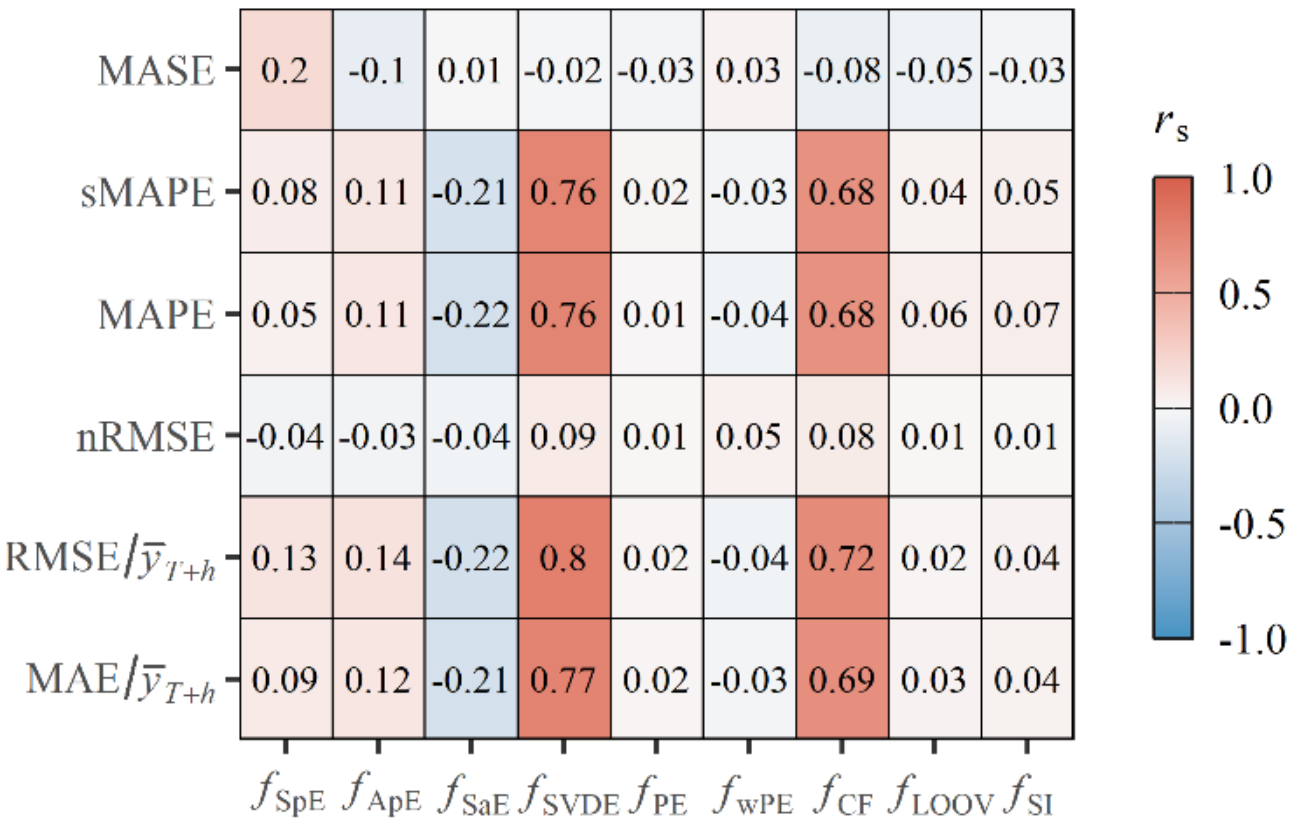

Fig. 8. Correlations between intrinsic predictability measures (time series features, columns) and realized predictability measures (forecasting accuracy metrics, rows); derived from STL-ARIMA forecasts on 2,807 time series using a two-year training set and a 36-week forecast horizon

### D. Application to PQ Monitoring

#### 1) Interpretation of SVD Entropy and Crest Factor

Fig. 9 presents a scatterplot of the two intrinsic features, $f_{\mathrm{SVDE}}$ and $f_{\mathrm{CF}}$, with points highlighted and colored according to their forecast accuracy (sMAPE), binned into three categories: good (0-10%), acceptable (10-25%) and poor (25-200%). The distribution reveals that most poor forecasts are associated with either high SVD entropy values (e.g. $f_{\mathrm{SVDE}} > 0.2$) and/or elevated Crest Factors (e.g. $f_{\mathrm{CF}} > 2$). Four examples are additionally highlighted and illustrated as TS in Fig. 10:

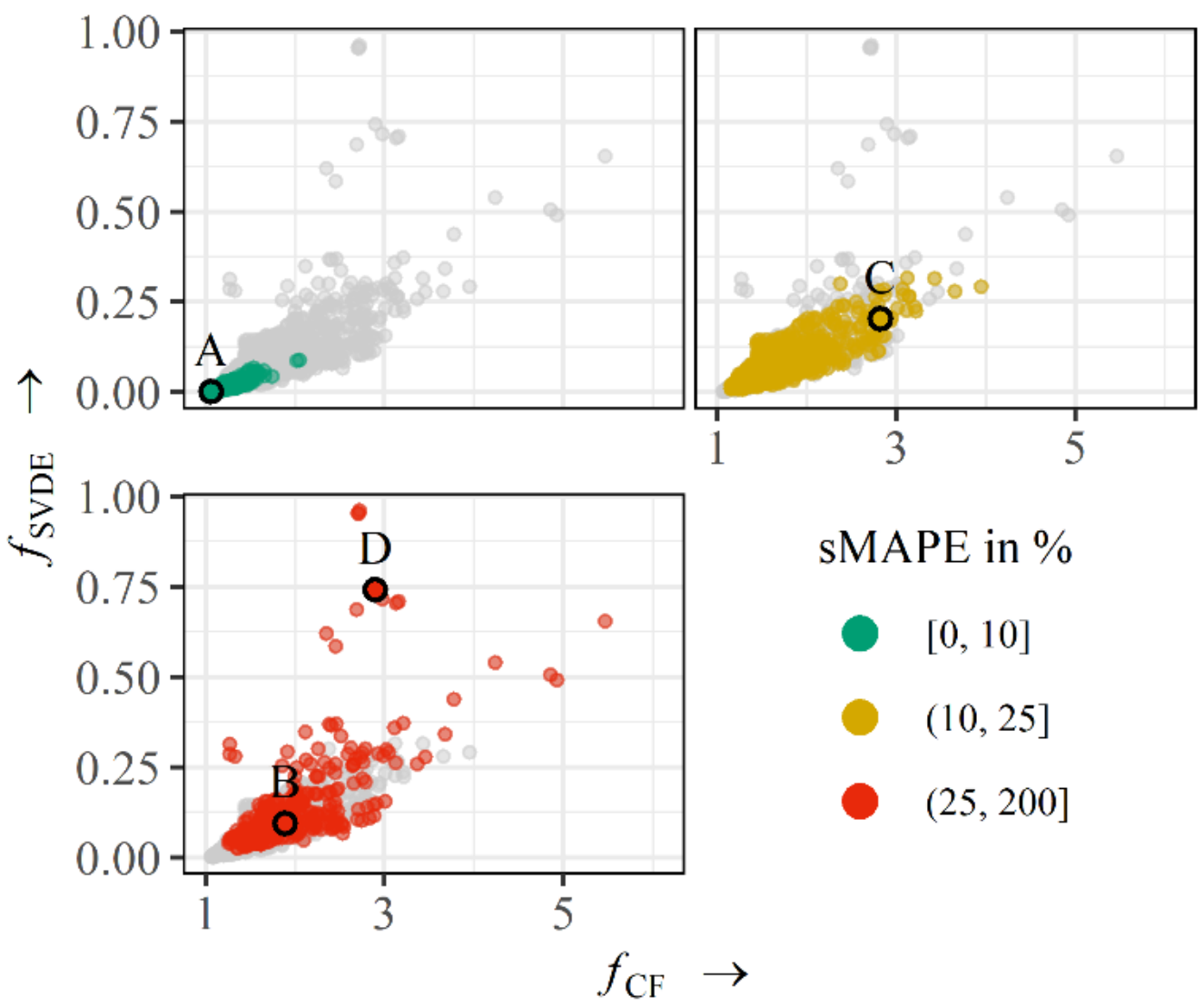

Fig. 9. Scatterplot of crest factor ($f_{\mathrm{CF}}$) and SVD entropy ($f_{\mathrm{SVDE}}$); sMAPE accuracy bins derived from STL-ARIMA forecasts on all 2,807 time series

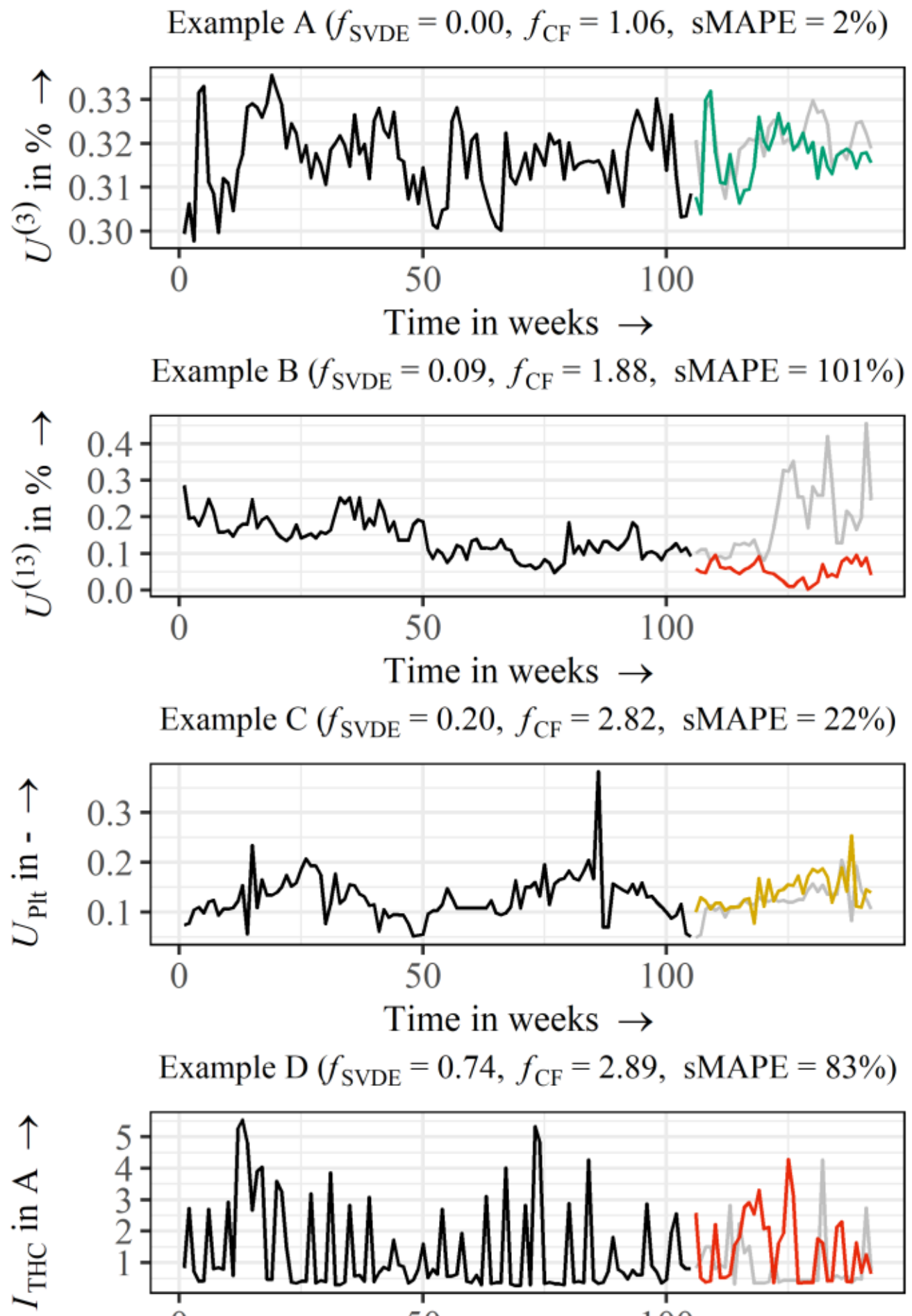

Fig. 10. Example time series from Fig. 9 with STL-ARIMA forecasts

Example A shows the 3[rd] harmonic voltage, which has low intrinsic features ($f_{\mathrm{SVDE}} = 0.00$, $f_{\mathrm{CF}} = 1.06$), indicating good predictability and resulting in a realized forecast accuracy of sMAPE = 2%.

Example B shows a case of deceptively good predictability based on low intrinsic features ($f_{\mathrm{SVDE}} = 0.09$, $f_{\mathrm{CF}} = 1.88$), but with very poor forecast accuracy (sMAPE = 101%). Here, a major structural change occurred within the test data that was not part of the training set. Such scenarios were frequently observed in higher order voltage harmonics. A large discrepancy between the intrinsically expected and the realized forecast accuracy may indicate an unforeseen structural change in the system, information that is itself valuable for network monitoring.

Example C shows higher intrinsic feature values ($f_{\mathrm{SVDE}} = 0.20$, $f_{\mathrm{CF}} = 2.82$), with the Crest Factor reflecting the visible spikes. Despite the elevated features suggesting poor predictability, the realized forecast accuracy remains acceptable (sMAPE = 22%), illustrating that high spikiness alone does not necessarily preclude satisfactory forecasts.

Example D exhibits very poor intrinsic predictability, indicated by a very high SVD entropy value ($f_{\mathrm{SVDE}} = 0.74$). The highly chaotic behavior results in a very poor forecast accuracy of sMAPE = 83%.

#### 2) Modeling the Probability of Poor Forecasts

To quantitatively assess how the intrinsic features relate to forecasting performance, the probability of obtaining a poor forecast was modeled directly. A Generalized Linear Model (GLM), specifically Logistic Regression, has been employed to estimate the probability that a forecast would be poor, defined as sMAPE > 25%. This probability was expressed as a function of the intrinsic features $f_{\mathrm{SVDE}}$ and $f_{\mathrm{CF}}$. A logistic function has been used to model this binary outcome:

$$\operatorname{logit}(P(\mathrm{sMAPE} > 25\%)) = \beta_0 + \beta_1 f_{\mathrm{SVDE}} + \beta_2 f_{\mathrm{CF}} \quad (13)$$

with

$$\operatorname{logit}(p) = \log\left(\frac{p}{1-p}\right)$$

The logit function ensures that the predicted probability remains within the interval $p \in [0,1]$ [47]. Fig. 11 shows the estimated probability of poor forecasts as a function of both intrinsic features.

Higher values of $f_{\mathrm{SVDE}}$ have a stronger impact on increasing the probability of poor forecasts, indicating reduced predictability. For $f_{\mathrm{SVDE}} > 0.2$ and $f_{\mathrm{CF}} > 2$, the probability of a poor forecast exceeds 60%. For $f_{\mathrm{SVDE}} > 0.3$, this probability exceeds 80% regardless of $f_{\mathrm{CF}}$.

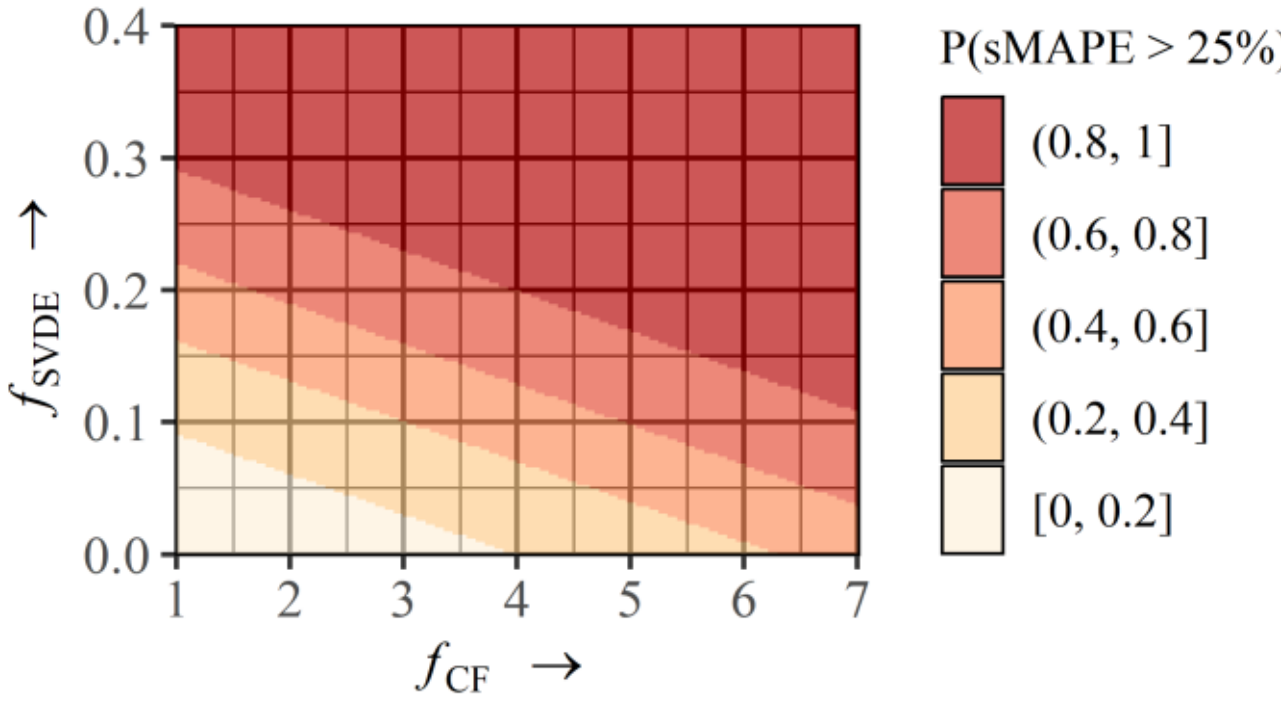

Fig. 11. Estimated probability of poor forecasts based on intrinsic features

## VI. Conclusion

Medium-term forecasting can extend the temporal scope of continuous PQ monitoring, but its value depends on whether its reliability can be judged for each time series individually. This study addressed that question in two stages, using 2,807 weekly time series from the German transmission system. Additive component models proved well suited: the STL-ARIMA hybrid achieved the highest accuracy, with an average sMAPE of 17.63% and a lower error than the seasonal naive benchmark for 72% of the time series. Accuracy depends strongly on the PQ parameter and the measurement site, with higher-order harmonics and long-term flicker remaining challenging for every model examined, which indicates that the limiting factor is often the time series itself rather than the choice of model.

Building on this, SVD entropy and the Crest Factor were introduced as intrinsic predictability measures. Both are computed from the training set alone, correlate strongly with the realized accuracy of all eight models, and allow the probability of a poor forecast to be estimated before any model is applied. This gives operators an a priori criterion for separating time series suitable for automated processing from those requiring manual review, primarily in network planning, where a parameter approaching its planning level calls for a forecast whose reliability is known beforehand. Although the analysis used transmission system data, the procedure makes no assumption specific to that voltage level.

Three directions are of interest for future work: probabilistic forecasts providing prediction intervals for risk assessment, accuracy measures that weight errors by their proximity to the applicable limit, and the use of the intrinsic measures for selecting the forecasting model rather than only for deciding whether to forecast at all.